\documentclass[fleqn]{llncs}
\usepackage{version}
\usepackage{jfis}

\title{On the Expressiveness of\\ Single-Pass Instruction Sequences%
       \thanks{This research was partly carried out in the framework of
               the  Jacquard-project Symbiosis, which is funded by the
               Netherlands Organisation for Scientific Research (NWO).}}
\author{J.A. Bergstra \and C.A. Middelburg }
\institute{Programming Research Group, University of Amsterdam, \\
           Kruislaan~403, 1098~SJ~Amsterdam, the Netherlands \\
           \email{J.A.Bergstra@uva.nl,C.A.Middelburg@uva.nl}}

\begin{document}
\maketitle

\begin{abstract}
We perceive programs as single-pass instruction sequences.
A single-pass instruction sequence under execution is considered to
produce a behaviour to be controlled by some execution environment.
Threads as considered in basic thread algebra model such behaviours.
We show that all regular threads, i.e.\ threads that can only be in a
finite number of states, can be produced by single-pass instruction
sequences without jump instructions if use can be made of Boolean
registers.
We also show that, in the case where goto instructions are used instead
of jump instructions, a bound to the number of labels restricts the
expressiveness.
\begin{classcode}
D.1.4, D.3.3, F.1.1, F.3.3.
\end{classcode}
\end{abstract}

\section{Introduction}
\label{sect-intro}

With the work presented in this paper, we carry on the line of research
with which a start was made in~\cite{BL02a}.
The working hypothesis of this line of research is that single-pass
instruction sequence is a central notion of computer science which
merits investigation for its own sake.
In this line of research, program algebra is taken for the basis of the
investigations.
Program algebra is a setting suited for investigating single-pass
instruction sequences.
It does not provide a notation for programs that is intended for actual
programming.

The starting-point of program algebra is the perception of a program as
a single-pass instruction sequence, i.e.\ a finite or infinite sequence
of instructions of which each instruction is executed at most once and
can be dropped after it has been executed or jumped over.
This perception is simple, appealing, and links up with practice.
A single-pass instruction sequence under execution is considered to
produce a behaviour to be controlled by some execution environment.
Threads as considered in basic thread algebra model such behaviours:
upon each action performed by a thread, a reply from the execution
environment determines how the thread proceeds.
A thread may make use of services, i.e.\ components of the execution
environment.

Each Turing machine can be simulated by means of a thread that makes use
of a service.
The thread and service correspond to the finite control and tape of the
Turing machine.
The threads that correspond to the finite controls of Turing machines
are examples of regular threads, i.e.\ threads that can only be in a
finite number of states.
The behaviours of all single-pass instruction sequences considered in
program algebra are regular threads and each regular thread is produced
by some single-pass instruction sequence.
In this paper, we show that each regular thread can be produced by some
single-pass instruction sequence without jump instructions if use can
be made of services that make up Boolean registers.

The primitive instructions of program algebra include jump instructions.
An interesting variant of program algebra is obtained by leaving out
jump instructions and adding labels and goto instructions.
It is easy to see that each regular thread can also be produced by some
single-pass instruction sequence with labels and goto instructions.
In this paper, we show that a bound to the number of labels restricts
the expressiveness of this variant.

This paper is organized as follows.
First, we review basic thread algebra and program algebra
(Sections~\ref{sect-BTA} and~\ref{sect-PGA}).
Next, we present a mechanism for interaction of threads with services
and give a description of Boolean register services
(Sections~\ref{sect-TSI} and~\ref{sect-Boolean-register}).
After that, we show that each regular thread can be produced by some
single-pass instruction sequence without jump instructions if use can be
made of Boolean register services (Section~\ref{sect-jump-free}).
Then, we introduce the variant of program algebra obtained by leaving out
jump instructions and adding labels and goto instructions
(Section~\ref{sect-PGAg}).
Following this, we show that a bound to the number of labels restricts
the expressiveness of this variant (Section~\ref{sect-labels}).
Finally, we make some concluding remarks (Section~\ref{sect-concl}).

\section{Basic Thread Algebra}
\label{sect-BTA}

In this section, we review \BTA, which is concerned with the behaviours
that sequential programs exhibit on execution.
These behaviours are called threads.

In \BTA, it is assumed that a fixed but arbitrary set $\BAct$ of
\emph{basic actions} has been given.
A thread performs actions in a sequential fashion.
Upon each action performed, a reply from the execution environment of
the thread determines how it proceeds.
To simplify matters, there are only two possible replies: $\True$ and
$\False$.

\BTA\ has one sort: the sort $\Thr$ of \emph{threads}.
To build terms of sort $\Thr$, it has the following constants and
operators:
\begin{iteml}
\item
the \emph{deadlock} constant $\const{\DeadEnd}{\Thr}$;
\item
the \emph{termination} constant $\const{\Stop}{\Thr}$;
\item
for each $a \in \BAct$, the binary \emph{postconditional composition}
operator $\funct{\pcc{\ph}{a}{\ph}}{\Thr \x \Thr}{\Thr}$.
\end{iteml}
We assume that there are infinitely many variables of sort $\Thr$,
including $x,y,z$.
We introduce \emph{action prefixing} as an abbreviation: $a \bapf p$
abbreviates $\pcc{p}{a}{p}$.

The thread denoted by a closed term of the form $\pcc{p}{a}{q}$ will
first perform $a$, and then proceed as the thread denoted by $p$
if the reply from the execution environment is $\True$ and proceed as
the thread denoted by $q$ if the reply from the execution environment is
$\False$.
The threads denoted by $\DeadEnd$ and $\Stop$ will become inactive and
terminate, respectively.
This implies that each closed \BTA\ term denotes a thread that will
become inactive or terminate after it has performed finitely many
actions.
Infinite threads can be described by guarded recursion.

A \emph{guarded recursive specification} over \BTA\ is a set of
recursion equations $E = \set{X = t_X \where X \in V}$, where $V$ is a
set of variables of sort $\Thr$ and each $t_X$ is a \BTA\ term of the
form $\DeadEnd$, $\Stop$ or $\pcc{t}{a}{t'}$ with $t$ and $t'$ that
contain only variables from $V$.
We write $\vars(E)$ for the set of all variables that occur in $E$.
We are only interested in models of \BTA\ in which guarded recursive
specifications have unique solutions, such as the projective limit model
of \BTA\ presented in~\cite{BB03a}.

For each guarded recursive specification $E$ and each $X \in \vars(E)$,
we introduce a constant $\rec{X}{E}$ of sort $\Thr$ standing for the
unique solution of $E$ for $X$.
The axioms for these constants are given in Table~\ref{axioms-REC}.%
\begin{table}[!t]
\caption{Axioms for guarded recursion}
\label{axioms-REC}
\begin{eqntbl}
\begin{saxcol}
\rec{X}{E} = \rec{t_X}{E} & \mif X \!=\! t_X \in E       & \axiom{RDP}
\\
E \Implies X = \rec{X}{E} & \mif X \in \vars(E)          & \axiom{RSP}
\end{saxcol}
\end{eqntbl}
\end{table}
In this table, we write $\rec{t_X}{E}$ for $t_X$ with, for all
$Y \in \vars(E)$, all occurrences of $Y$ in $t_X$ replaced by
$\rec{Y}{E}$.\linebreak[2]
$X$, $t_X$ and $E$ stand for an arbitrary variable of sort $\Thr$, an
arbitrary \BTA\ term of sort $\Thr$ and an arbitrary guarded recursive
specification over \BTA, respectively.
Side conditions are added to restrict what $X$, $t_X$ and $E$ stand for.

Closed terms that denote the same infinite thread cannot always be
proved equal by means of the axioms given in Table~\ref{axioms-REC}.
We introduce \AIP\ (Approximation Induction Principle) to remedy this.
\AIP\ is based on the view that two threads are identical if their
approximations up to any finite depth are identical.
The approximation up to depth $n$ of a thread is obtained by cutting it
off after it has performed $n$ actions.
In \AIP, the approximation up to depth $n$ is phrased in terms of the
unary \emph{projection} operator $\funct{\projop{n}}{\Thr}{\Thr}$.
\AIP\ and the axioms for the projection operators are given in
Table~\ref{axioms-AIP}.%
\begin{table}[!t]
\caption{Approximation induction principle}
\label{axioms-AIP}
\begin{eqntbl}
\begin{axcol}
\AND{n \geq 0} \proj{n}{x} = \proj{n}{y} \Implies x = y & \axiom{AIP} \\
\proj{0}{x} = \DeadEnd                                  & \axiom{P0} \\
\proj{n+1}{\Stop} = \Stop                               & \axiom{P1} \\
\proj{n+1}{\DeadEnd} = \DeadEnd                         & \axiom{P2} \\
\proj{n+1}{\pcc{x}{a}{y}} =
                      \pcc{\proj{n}{x}}{a}{\proj{n}{y}} & \axiom{P3}
\end{axcol}
\end{eqntbl}
\end{table}

\section{Program Algebra}
\label{sect-PGA}

In this section, we review \PGA.
The perception of a program as a single-pass instruction sequence is the
starting-point of \PGA.

In \PGA, it is assumed that a fixed but arbitrary set $\BInstr$ of
\emph{basic instructions} has been given.
\PGA\ has the following \emph{primitive instructions}:
\begin{iteml}
\item
for each $a \in \BInstr$, a \emph{plain basic instruction} $a$;
\item
for each $a \in \BInstr$, a \emph{positive test instruction} $\ptst{a}$;
\item
for each $a \in \BInstr$, a \emph{negative test instruction} $\ntst{a}$;
\item
for each $l \in \Nat$, a \emph{forward jump instruction} $\fjmp{l}$;
\item
a \emph{termination instruction} $\halt$.
\end{iteml}
We write $\PInstr$ for the set of all primitive instructions.

The intuition is that the execution of a basic instruction $a$ produces
either $\True$ or $\False$ at its completion.
In the case of a positive test instruction $\ptst{a}$, $a$ is executed
and execution proceeds with the next primitive instruction if $\True$ is
produced.
Otherwise, the next primitive instruction is skipped and execution
proceeds with the primitive instruction following the skipped one.
If there is no next instruction to be executed, deadlock occurs.
In the case of a negative test instruction $\ntst{a}$, the role of the
value produced is reversed.
In the case of a plain basic instruction $a$, execution always proceeds
as if $\True$ is produced.
The effect of a forward jump instruction $\fjmp{l}$ is that execution
proceeds with the $l$-th next instruction.
If $l$ equals $0$ or the $l$-th next instruction does not exist,
deadlock occurs.
The effect of the termination instruction $\halt$ is that execution
terminates.

\PGA\ has the following constants and operators:
\begin{iteml}
\item
for each $u \in \PInstr$, an \emph{instruction} constant $u$\,;
\item
the binary \emph{concatenation} operator $\ph \conc \ph$\,;
\item
the unary \emph{repetition} operator $\ph\rep$\,.
\end{iteml}
We assume that there are infinite many variables, including $X,Y,Z$.

A closed \PGA\ term is considered to denote a non-empty, finite or
periodic infinite sequence of primitive instructions.%
\footnote
{A periodic infinite sequence is an infinite sequence with only finitely
 many subsequences.}
Closed \PGA\ terms are considered equal if they denote the same
instruction sequence.
The axioms for instruction sequence equivalence are given in
Table~\ref{axioms-PGA}.%
\begin{table}[!t]
\caption{Axioms of \PGA}
\label{axioms-PGA}
\begin{eqntbl}
\begin{axcol}
(X \conc Y) \conc Z = X \conc (Y \conc Z)              & \axiom{PGA1} \\
(X^n)\rep = X\rep                                      & \axiom{PGA2} \\
X\rep \conc Y = X\rep                                  & \axiom{PGA3} \\
(X \conc Y)\rep = X \conc (Y \conc X)\rep              & \axiom{PGA4}
\end{axcol}
\end{eqntbl}
\end{table}
In this table, $n$ stands for an arbitrary natural number greater than
$0$.
For each \PGA\ term $P$, the term $P^n$ is defined by induction on $n$
as follows: $P^1 = P$ and $P^{n+1} = P \conc P^n$.
The equation $X\rep = X \conc X\rep$ is derivable.
Each closed \PGA\ term is derivably equal to one of the form $P$ or
$P \conc Q\rep$, where $P$ and $Q$ are closed \PGA\ terms in which the
repetition operator does not occur.

The behaviours of the instruction sequences denoted by closed \PGA\
terms are considered threads, with basic instructions taken for basic
actions.
The \emph{thread extraction} operation $\extr{\ph}$ determines, for each
closed \PGA\ term $P$, a closed term of \BTA\ with guarded recursion
that denotes the behaviour of the instruction sequence denoted by $P$.
The thread extraction operation is defined by the equations given in
Table~\ref{axioms-thread-extr} (for $a \in \BInstr$, $l \in \Nat$ and
$u \in \PInstr$)%
\begin{table}[!t]
\caption{Defining equations for thread extraction operation}
\label{axioms-thread-extr}
\begin{eqntbl}
\begin{eqncol}
\extr{a} = a \bapf \DeadEnd \\
\extr{a \conc X} = a \bapf \extr{X} \\
\extr{\ptst{a}} = a \bapf \DeadEnd \\
\extr{\ptst{a} \conc X} =
\pcc{\extr{X}}{a}{\extr{\fjmp{2} \conc X}} \\
\extr{\ntst{a}} = a \bapf \DeadEnd \\
\extr{\ntst{a} \conc X} =
\pcc{\extr{\fjmp{2} \conc X}}{a}{\extr{X}}
\end{eqncol}
\qquad
\begin{eqncol}
\extr{\fjmp{l}} = \DeadEnd \\
\extr{\fjmp{0} \conc X} = \DeadEnd \\
\extr{\fjmp{1} \conc X} = \extr{X} \\
\extr{\fjmp{l+2} \conc u} = \DeadEnd \\
\extr{\fjmp{l+2} \conc u \conc X} = \extr{\fjmp{l+1} \conc X} \\
\extr{\halt} = \Stop \\
\extr{\halt \conc X} = \Stop
\end{eqncol}
\end{eqntbl}
\end{table}
and the rule that $\extr{\fjmp{l} \conc X} = \DeadEnd$ if $\fjmp{l}$ is
the beginning of an infinite jump chain.
This rule is formalized in e.g.~\cite{BM07g}.

\section{Interaction of Threads with Services}
\label{sect-TSI}

A thread may make use of services.
That is, a thread may perform an action for the purpose of interacting
with a service that takes the action as a command to be processed.
The processing of an action may involve a change of state of the service
and at completion of the processing of the action the service returns a
reply value to the thread.
In this section, we introduce the use operators, which are concerned with
this kind of interaction between threads and services.

It is assumed that a fixed but arbitrary set $\Foci$ of \emph{foci} and
a fixed but arbitrary set $\Meth$ of \emph{methods} have been given.
Each focus plays the role of a name of some service provided by an
execution environment that can be requested to process a command.
Each method plays the role of a command proper.
For the set $\BAct$ of actions, we take the set
$\set{f.m \where f \in \Foci, m \in \Meth}$.
Performing an action $f.m$ is taken as making a request to the
service named $f$ to process command~$m$.

A \emph{service} $H$ consists of
\begin{iteml}
\item
a set $S$ of \emph{states};
\item
an \emph{effect} function $\funct{\eff}{\Meth \x S}{S}$;
\item
a \emph{yield} function
$\funct{\yld}{\Meth \x S}{\set{\True,\False,\Blocked}}$;
\item
an \emph{initial state} $s_0 \in S$;
\end{iteml}
satisfying the following condition:
\begin{ldispl}
\Forall{m \in \Meth, s \in S}
{(\yld(m,s) = \Blocked \Implies
  \Forall{m' \in \Meth}{\yld(m',\eff(m,s)) = \Blocked})}\;.
\end{ldispl}
The set $S$ contains the states in which the service may be, and the
functions $\eff$ and $\yld$ give, for each method $m$ and state $s$, the
state and reply, respectively, that result from processing $m$ in state
$s$.

Let $H = \tup{S,\eff,\yld,s_0}$ be  a service and let $m \in \Meth$.
Then
the \emph{derived service} of $H$ after processing $m$, written
$\derive{m}H$, is the service $\tup{S,\eff,\yld,\eff(m,s_0)}$; and
the \emph{reply} of $H$ after processing $m$, written $H(m)$, is
$\yld(m,s_0)$.

When a thread makes a request to service $H$ to process $m$:
\begin{iteml}
\item
if $H(m) \neq \Blocked$, then the request is accepted, the reply is
$H(m)$, and the service proceeds as $\derive{m}H$;
\item
if $H(m) = \Blocked$, then the request is rejected.
\end{iteml}

We introduce the sort $\Serv$ of \emph{services} and,
for each $f \in \Foci$, the binary \emph{use} operator
$\funct{\use{\ph}{f}{\ph}}{\Thr \x \Serv}{\Thr}$.
The axioms for these operators are given in Table~\ref{axioms-TSU}.%
\begin{table}[!t]
\caption{Axioms for use operators}
\label{axioms-TSU}
\begin{eqntbl}
\begin{saxcol}
\use{\Stop}{f}{H} = \Stop                            & & \axiom{TSU1} \\
\use{\DeadEnd}{f}{H} = \DeadEnd                      & & \axiom{TSU2} \\
\use{(\pcc{x}{g.m}{y})}{f}{H} =
\pcc{(\use{x}{f}{H})}{g.m}{(\use{y}{f}{H})}
 & \mif f \neq g                                       & \axiom{TSU3} \\
\use{(\pcc{x}{f.m}{y})}{f}{H} = \use{x}{f}{\derive{m}H}
                                & \mif H(m) = \True    & \axiom{TSU4} \\
\use{(\pcc{x}{f.m}{y})}{f}{H} = \use{y}{f}{\derive{m}H}
                                & \mif H(m) = \False   & \axiom{TSU5} \\
\use{(\pcc{x}{f.m}{y})}{f}{H} = \DeadEnd
                                & \mif H(m) = \Blocked & \axiom{TSU6} \\
\AND{n \geq 0} \use{\proj{n}{x}}{f}{H} = \use{\proj{n}{y}}{f}{H}
 \Implies \use{x}{f}{H} = \use{y}{f}{H}              & & \axiom{TSU7}
\end{saxcol}
\end{eqntbl}
\end{table}

Intuitively, $\use{p}{f}{H}$ is the thread that results from processing
all actions performed by thread $p$ that are of the form $f.m$ by
service $H$.
When an action of the form $f.m$ performed by thread $p$ is processed by
service $H$, the postconditional composition concerned is eliminated on
the basis of the reply value produced.
No internal action is left as a trace of the processed action, like
with the use operators found in papers on thread interleaving (see
e.g.~\cite{BM07a}).

Combining TSU2 and TSU7, we obtain
$\AND{n \geq 0} \use{\proj{n}{x}}{f}{H} = \DeadEnd \Implies
 \use{x}{f}{H} = \DeadEnd$.

\section{Instruction Sequences Acting on Boolean Registers}
\label{sect-Boolean-register}

Our study of jump-free instruction sequences in
Section~\ref{sect-jump-free} is concerned with instruction sequences
that act on Boolean registers.
In this section, we describe services that make up Boolean registers.

A Boolean register service accepts the following methods:
\begin{itemize}
\item
a \emph{set to true method} $\setbr{\True}$;
\item
a \emph{set to false method} $\setbr{\False}$;
\item
a \emph{get method} $\getbr$.
\end{itemize}
We write $\Methbr$ for the set
$\set{\setbr{\True},\setbr{\False},\getbr}$.
It is assumed that $\Methbr \subseteq \Meth$.

The methods accepted by Boolean register services can be explained as
follows:
\begin{itemize}
\item
$\setbr{\True}$\,:
the contents of the Boolean register becomes $\True$ and the reply is
$\True$;
\item
$\setbr{\False}$\,:
the contents of the Boolean register becomes $\False$ and the reply is
$\False$;
\item
$\getbr$\,:
nothing changes and the reply is the contents of the Boolean register.
\end{itemize}

Let $s \in \set{\True,\False,\Blocked}$.
Then the \emph{Boolean register service} with initial state $s$, written
$\BR_s$, is the service $\tup{\set{\True,\False,\Blocked},\eff,\eff,s}$,
where the function $\eff$ is defined as follows
($b \in \set{\True,\False}$):
\begin{ldispl}
\begin{geqns}
\eff(\setbr{\True},b) = \True\;,\;
\\
\eff(\setbr{\False},b) = \False\;,
\\
\eff(\getbr,b) = b\;,
\end{geqns}
\qquad\qquad
\begin{geqns}
\eff(m,b) = \Blocked & \mif m \not\in \Methbr\;,
\\
\eff(m,\Blocked) = \Blocked\;.
\end{geqns}
\end{ldispl}
Notice that the effect and yield functions of a Boolean register service
are the same.

\section{Jump-Free Instruction Sequences}
\label{sect-jump-free}

In this section, we show that each thread that can only be in a finite
number of states can be produced by some single-pass instruction
sequence without jump instructions if use can be made of Boolean
register services.

First, we make precise what it means that a thread can only be in a
finite number of states.
We assume that a fixed but arbitrary model $\fM$ of \BTA\ extended with
guarded recursion and the use mechanism has been given, we use the term
thread only for the elements from the domain of $\fM$, and we denote the
interpretations of constants and operators in $\fM$ by the constants and
operators themselves.

Let $p$ be a thread.
Then the set of \emph{states} or \emph{residual threads} of $p$,
written $\Res(p)$, is inductively defined as follows:
\begin{itemize}
\item
$p \in \Res(p)$;
\item
if $\pcc{q}{a}{r} \in \Res(p)$, then $q \in \Res(p)$ and
$r \in \Res(p)$.
\end{itemize}
We say that $p$ is a \emph{regular} thread if $\Res(p)$ is finite.

We will make use of the fact that being a regular thread
coincides with being the solution of a finite guarded recursive
specification of a restricted form.

A \emph{linear recursive specification} over \BTA\ is a guarded
recursive specification $E = \set{X = t_X \where X \in V}$, where each
$t_X$ is a term of the form $\DeadEnd$, $\Stop$ or $\pcc{Y}{a}{Z}$ with
$Y,Z \in V$.
\begin{proposition}
\label{prop-lin-rec}
Let $p$ be a thread.
Then $p$ is a regular thread iff there exists a finite linear recursive
specification $E$ and a variable $X \in \vars(E)$ such that $p$ is the
solution of $E$ for $X$.
\end{proposition}
\begin{proof}
This proposition generalizes Theorem~1 from~\cite{PZ06a} from the
projective limit model to an arbitrary model.
However, the proof of that theorem is applicable to any model.
\qed
\end{proof}

In the proof of the next theorem, we associate a closed \PGA\ term $P$
in which jump instructions do not occur with a finite linear recursive
specification
\begin{ldispl}
E = \set{X_i = \pcc{X_{l(i)}}{a_i}{X_{r(i)}} \where i \in [1,n]} \union
\set{X_{n+1} = \Stop, X_{n+2} = \DeadEnd}\;.
\end{ldispl}
In $P$, a number of Boolean register services is used for specific
purposes.
The purpose of each individual Boolean register is reflected in the
focus that serves as its name:
\begin{iteml}
\item
for each $i \in [1,n + 2]$, $\stbr{i}$ serves as the name of a Boolean
register that is used to indicate whether the current state of
$\rec{X_1}{E}$ is $\rec{X_i}{E}$;
\item
$\rtbr$ serves as the name of a Boolean register that is used to
indicate whether the reply upon the action performed by $\rec{X_1}{E}$
in its current state is $\True$;
\item
$\rfbr$ serves as the name of a Boolean register that is used to
indicate whether the reply upon the action performed by $\rec{X_1}{E}$
in its current state is $\False$;
\item
$\enbr$ serves as the name of a Boolean register that is used to achieve
that instructions not related to the current state of $\rec{X_1}{E}$ are
passed correctly;
\item
$\skbr$ serves as the name of a Boolean register that is used to achieve
with the instruction $\ptst{\skbr.\setbr{\False}}$ that the following
instruction is skipped.
\end{iteml}
Now we turn to the theorem announced above.
It states rigorously that the solution of every finite linear recursive
specification can be produced by an instruction sequence without jump
instructions if use can be made of Boolean register services.
\begin{theorem}
\label{theorem-jump-free}
Let a finite linear recursive specification
\begin{ldispl}
E = \set{X_i = \pcc{X_{l(i)}}{a_i}{X_{r(i)}} \where i \in [1,n]} \union
\set{X_{n+1} = \Stop, X_{n+2} = \DeadEnd}
\end{ldispl}
be given.
Then there exists a closed \PGA\ term $P$ in which jump instructions do
not occur such that
\begin{ldispl}
 \rec{X_1}{E}
\\ \;\; {} =
 (((( \ldots
     (\extr{P}
       \useop{\stbr{1}} \BR_\False) \ldots
       \useop{\stbr{n{+}2}} \BR_\False)
     \useop{\rtbr} \BR_\False) \useop{\rfbr} \BR_\False)
   \useop{\enbr} \BR_\False) \useop{\skbr} \BR_\False\;.
\end{ldispl}
\end{theorem}
\begin{proof}
We associate a closed \PGA\ term $P$ in which jump instructions do not
occur with $E$ as follows:
\begin{ldispl}
P =
\stbr{1}.\setbr{\True} \conc (Q_1 \conc \ldots \conc Q_{n+1})\rep\;,
\end{ldispl}
where, for each $i \in [1,n]$:
\begin{ldispl}
\begin{aeqns}
Q_i & = &
\ptst{\stbr{i}.\getbr} \conc \enbr.\setbr{\True} \conc {}
\\ \phantom{Q_{n+1}} & &
\ptst{\stbr{i}.\getbr} \conc \stbr{i}.\setbr{\False} \conc {}
\\ & &
\ptst{\enbr.\getbr} \conc \ntst{a_i} \conc
\ptst{\skbr.\setbr{\False}} \conc \rtbr.\setbr{\True} \conc {}
\\ & &
\ptst{\enbr.\getbr} \conc \ptst{\rtbr.\getbr} \conc
\ptst{\skbr.\setbr{\False}} \conc \rfbr.\setbr{\True} \conc {}
\\ & &
\ptst{\rtbr.\getbr} \conc \stbr{l(i)}.\setbr{\True} \conc {}
\\ & &
\ptst{\rfbr.\getbr} \conc \stbr{r(i)}.\setbr{\True} \conc {}
\\ & &
\rtbr.\setbr{\False} \conc \rfbr.\setbr{\False} \conc
\enbr.\setbr{\False}\;,
\end{aeqns}
\end{ldispl}
and
\begin{ldispl}
Q_{n+1} = \ptst{\stbr{n{+}1}.\getbr} \conc \halt\;.
\end{ldispl}
We use the following abbreviations
(for $i \in [1,n + 1]$ and $j \in [1,n + 2]$):
\begin{trivlist}
\item
$P'_i\phantom{\extrbr{}{j}}$ for
$Q_i \conc \ldots \conc Q_{n+1} \conc
 (Q_1 \conc \ldots \conc Q_{n+1})\rep$;
\item
$\extrbr{P'_i}{j}$ for
$(((( \ldots
     (\extr{P'_i}
       \useop{\stbr{1}} \BR_{b_1}) \ldots
       \useop{\stbr{n{+}2}} \BR_{b_{n+2}})
     \useop{\rtbr} \BR_\False) \useop{\rfbr} \BR_\False)
   \useop{\enbr} \BR_\False) \useop{\skbr} \BR_\False$,
where $b_j = \True$ and, for each $j' \in [1,n + 2]$ such that
$j' \neq j$, $b_{j'} = \False$.
\end{trivlist}
From the definition of thread extraction, the definition of Boolean
register services, and axiom TSU4, it follows that
\begin{ldispl}
 (((( \ldots
     (\extr{P}
       \useop{\stbr{1}} \BR_\False) \ldots
       \useop{\stbr{n{+}2}} \BR_\False)
     \useop{\rtbr} \BR_\False) \useop{\rfbr} \BR_\False)
   \useop{\enbr} \BR_\False) \useop{\skbr} \BR_\False
\\ \;\; {} =
 \extrbr{P'_1}{1}\;.
\end{ldispl}
This leaves us to show that $\rec{X_1}{E} = \extrbr{P'_1}{1}$.

Using the definition of thread extraction, the definition of Boolean
register services, and axioms P0, P2, TSU1, TSU2, TSU4, TSU5 and TSU7,
we easily prove the following:
\begin{ldispl}
\begin{aceqns}
\extrbr{P'_i}{j} & = & \extrbr{P'_{i+1}}{j}
 & \mif 1 \leq i \leq n \And 1 \leq j \leq n + 1 \And i \neq j
                                                           \hsp{.49} (1)
\\
\extrbr{P'_i}{j} & = & \extrbr{P'_1}{j}
 & \mif i = n + 1 \And 1 \leq j \leq n + 1 \And i \neq j      \hfill (2)
\\
\extrbr{P'_i}{i} & = &
\pcc{\extrbr{P'_{i+1}}{l(i)}}{a_i}{\extrbr{P'_{i+1}}{r(i)}}
 & \mif 1 \leq i \leq n                                       \hfill (3)
\\
\extrbr{P'_i}{i} & = & \Stop & \mif i = n + 1                 \hfill (4)
\\
\extrbr{P'_i}{j} & = & \DeadEnd
 & \mif 1 \leq i \leq n + 1 \And j = n + 2                    \hfill (5)
\end{aceqns}
\end{ldispl}
From Properties~1 and~2, it follows that
\begin{ldispl}
\begin{aceqns}
\extrbr{P'_i}{j} & = & \extrbr{P'_{j}}{j}
 & \mif 1 \leq i \leq n + 1 \And 1 \leq j \leq n + 1 \And i \neq j\;.
\end{aceqns}
\end{ldispl}
From this and Property~3, it follows that
\begin{ldispl}
\begin{aceqns}
\extrbr{P'_i}{i} & = &
\pcc{\extrbr{P'_{l(i)}}{l(i)}}{a_i}{\extrbr{P'_{r(i)}}{r(i)}}
 & \mif 1 \leq i \leq n\;.
\end{aceqns}
\end{ldispl}
From this and Properties~4 and~5, it follows that $\extrbr{P'_1}{1}$ is
a solution of $E$ for $X_1$.
Because linear recursive specifications have unique solutions, it
follows that $\rec{X_1}{E} = \extrbr{P'_1}{1}$.
\qed
\end{proof}

Theorem~\ref{theorem-jump-free} goes through in the case where
$E = \set{X_1 = \DeadEnd}$: a witnessing $P$ is $(\skbr.\getbr)\rep$.
It follows from the proof of Proposition~\ref{prop-lin-rec} given
in~\cite{PZ06a} that, for each regular thread $p$, either $p$ is the
solution of $\set{X_1 = \DeadEnd}$ for $X_1$ or there exists a finite
linear recursive specification $E$ of the form considered in
Theorem~\ref{theorem-jump-free} such that $p$ is the solution of $E$ for
$X_1$.
Hence, we have the following corollary of Proposition~\ref{prop-lin-rec}
and Theorem~\ref{theorem-jump-free}:
\begin{corollary}
\label{corollary-jump-free}
For each regular thread $p$, there exists a closed \PGA\ term $P$ in
which jump instructions do not occur such that $p$ is the thread denoted
by
\begin{ldispl}
 (((( \ldots
     (\extr{P}
       \useop{\stbr{1}} \BR_\False) \ldots
       \useop{\stbr{n{+}2}} \BR_\False)
     \useop{\rtbr} \BR_\False) \useop{\rfbr} \BR_\False)
   \useop{\enbr} \BR_\False) \useop{\skbr} \BR_\False\;.
\end{ldispl}
\end{corollary}
In other words, each regular thread can be produced by an instruction
sequence without jump instructions if use can be made of Boolean
register services.

The construction of such instructions sequences given in the proof of
Theorem~\ref{theorem-jump-free} is weakly reminiscent of the
construction of structured programs from flow charts found
in~\cite{Coo67a}.
However, our construction is more extreme: it yields programs that
contain neither unstructured jumps nor a rendering of the conditional
and loop constructs used in structured programming.

\section{Program Algebra with Labels and Goto's}
\label{sect-PGAg}

In this section, we introduce \PGAg, a variant of \PGA\ obtained by
leaving out jump instructions and adding labels and goto instructions.

In \PGAg, like in \PGA, it is assumed that a fixed but arbitrary set
$\BInstr$ of basic instructions has been given.
\PGAg\ has the following \emph{primitive instructions}:
\begin{iteml}
\item
for each $a \in \BInstr$, a \emph{plain basic instruction} $a$;
\item
for each $a \in \BInstr$, a \emph{positive test instruction} $\ptst{a}$;
\item
for each $a \in \BInstr$, a \emph{negative test instruction} $\ntst{a}$;
\item
for each $l \in \Nat$, a \emph{label instruction} $\lbl{l}$;
\item
for each $l \in \Nat$, a \emph{goto instruction} $\goto{l}$;
\item
a \emph{termination instruction} $\halt$.
\end{iteml}
We write $\PInstrg$ for the set of all primitive instructions of \PGAg.

The plain basic instructions, the positive test instructions, the
negative test instructions, and the termination instruction are as in
\PGA.
Upon execution, a label instruction $\lbl{l}$ is simply skipped.
If there is no next instruction to be executed, deadlock occurs.
The effect of a goto instruction $\goto{l}$ is that execution
proceeds with the occurrence of the label instruction $\lbl{l}$ next
following if it exists.
If there is no occurrence of the label instruction $\lbl{l}$, deadlock
occurs.

\PGAg\ has a constant $u$ for each $u \in \PInstrg$.
The operators of \PGAg\ are the same as the operators as \PGA.
Likewise, the axioms of \PGAg\ are the same as the axioms as \PGA.

Just like in the case of \PGA, the behaviours of the instruction
sequences denoted by closed \PGAg\ terms are considered threads.
The behaviours of the instruction sequences denoted by closed \PGAg\
terms are indirectly given by the behaviour preserving
function $\pgagpga$ from the set of all closed \PGAg\ terms to the set
of all closed \PGA\ terms defined by
\begin{ldispl}
\pgagpga(u_1 \conc \ldots \conc u_n) =
\pgagpga(u_1 \conc \ldots \conc u_n \conc (\goto{1})\rep)\;,
\eqnsep
\pgagpga(u_1 \conc \ldots \conc u_n \conc
         (u_{n+1} \conc \ldots \conc u_m)\rep)
\\ \;\; {} =
\phi_1(u_1) \conc \ldots \conc \phi_n(u_n) \conc
(\phi_{n+1}(u_{n+1}) \conc \ldots \conc \phi_m(u_m))\rep\;,
\end{ldispl}
where the auxiliary functions $\funct{\phi_j}{\PInstrg}{\PInstr}$ are
defined as follows ($1 \leq j \leq m$):
\begin{ldispl}
\begin{aceqns}
\phi_j(\lbl{l})  & = & \fjmp{1}\;,                                  & \\
\phi_j(\goto{l}) & = & \fjmp{\tgt_j(l)}\;,                          & \\
\phi_j(u)        & = & u
 & \mif u\; \mathrm{is\;not\;a\;label\; or\;goto\;instruction}\;,
\end{aceqns}
\end{ldispl}
where
\begin{iteml}
\item
$\tgt_j(l) = i$ if the leftmost occurrence of $\lbl{l}$ in
$u_j \conc \ldots \conc u_m \conc u_{n+1} \conc \ldots \conc u_m$ is the
$i$-th instruction;
\item
$\tgt_j(l) = 0$ if there are no occurrences of $\lbl{l}$ in
$u_j \conc \ldots \conc u_m \conc u_{n+1} \conc \ldots \conc u_m$.
\end{iteml}
Let $P$ be a closed \PGAg\ term.
Then the behaviour of $P$ is $\extr{\pgagpga(P)}$.
The approach to semantics followed here is introduced under the name
\emph{projection semantics} in~\cite{BL02a}.
The function $\pgagpga$ is called a \emph{projection}.

\section{A Bounded Number of Labels}
\label{sect-labels}

In this section, we show that a bound to the number of labels restricts
the expressiveness of \PGAg.
We will refer to \PGAg\ terms that do not contain label instructions
$\lbl{l}$ with $l > k$ as \bPGAg{k} terms.
Moreover, we will write $\bPInstrg{k}$ for the set
$\PInstrg \diff \set{\lbl{l} \where l > k}$.

We define an alternative projection for closed \bPGAg{k} terms, which
takes into account that these terms contain only label instructions
$\lbl{l}$ with $1 \leq l \leq k$.
The alternative projection $\bpgagpga{k}$ from the set of all closed
\bPGAg{k} terms to the set of all closed \PGA\ terms is defined by
\begin{ldispl}
\bpgagpga{k}(u_1 \conc \ldots \conc u_n) =
\bpgagpga{k}(u_1 \conc \ldots \conc u_n \conc (\goto{1})\rep)\;,
\eqnsep
\bpgagpga{k}(u_1 \conc \ldots \conc u_n \conc
            (u_{n+1} \conc \ldots \conc u_m)\rep)
\\ \;\; {} =
\psi(u_1,u_2) \conc \ldots \conc \psi(u_n,u_{n+1}) \conc {}
\\ \;\; \phantom{{} = {}}
(\psi(u_{n+1},u_{n+2}) \conc \ldots \conc
 \psi(u_{m-1},u_m) \conc \psi(u_m,u_{n+1}))\rep\;,
\end{ldispl}
where the auxiliary function
$\funct{\psi}{\bPInstrg{k} \x \bPInstrg{k}}{\PInstr}$
is defined as follows:
\begin{ldispl}
\begin{aeqns}
\psi(u',u'')  & = &
\psi'(u') \conc \fjmp{k{+}2} \conc \fjmp{k{+}2} \conc \psi''(u'')\;,
\end{aeqns}
\end{ldispl}
where the auxiliary functions
$\funct{\psi',\psi''}{\bPInstrg{k}}{\PInstr}$
are defined as follows:
\begin{ldispl}
\begin{aceqns}
\psi'(\lbl{l})  & = & \fjmp{1}\;,                                   & \\
\psi'(\goto{l}) & = & \fjmp{l{+}2} & \mif l \leq k\;,                 \\
\psi'(\goto{l}) & = & \fjmp{0}     & \mif l > k\;,                    \\
\psi'(u)        & = & u
 & \mif u\; \mathrm{is\;not\;a\;label\; or\;goto\;instruction}\;,
\eqnsep
\psi''(\lbl{l}) & = &
\multicolumn{2}{@{}l@{}}
 {(\fjmp{k{+}3})^{l-1} \conc \fjmp{k{-}l{+}1} \conc
  (\fjmp{k{+}3})^{k-l}\;,}  \\
\psi''(u)       & = & (\fjmp{k{+}3})^k
 & \mif u\; \mathrm{is\;not\;a\;label\;instruction}\;.
\end{aceqns}
\end{ldispl}

In order to clarify the alternative projection, we explain how the
intended effect of a goto instruction is obtained.
If $u_j$ is $\goto{l}$, then $\psi'(u_j)$ is $\fjmp{l{+}2}$.
The effect of $\fjmp{l{+}2}$ is a jump to the $l$-th instruction in
$\psi''(u_{j+1})$ if $j < m$ and a jump to the $l$-th instruction in
$\psi''(u_{n+1})$ if $j = m$.
If this instruction is $\fjmp{k{-}l{+}1}$, then its effect is a jump to
the occurrence of $\fjmp{1}$ that replaces $\lbl{l}$.
However, if this instruction is $\fjmp{k{+}3}$, then its effect is a
jump to the $l$-th instruction in $\psi''(u_{j+2})$ if $j < m - 1$, a
jump to the $l$-th instruction in $\psi''(u_{n+1})$ if $j = m - 1$, and
a jump to the $l$-th instruction in $\psi''(u_{n+2})$ if $j = m$.

In the proof of Theorem~\ref{theorem-labels} below, chains of forward
jumps are removed in favour of single jumps.
The following proposition justifies these removals.
\begin{proposition}
\label{prop-jump-chains}
For each \PGA\ context $C[\,]$:
\begin{ldispl}
\extr{C[\fjmp{n+1} \conc u_1 \conc \ldots \conc u_n \conc \fjmp{m}]} =
\extr{C[\fjmp{m+n+1} \conc u_1 \conc \ldots \conc u_n \conc \fjmp{m}]}
\;.
\end{ldispl}
\end{proposition}
\begin{proof}
Contexts of the forms $C[\,]\rep \conc Q$ and
$P \conc C[\,]\rep \conc Q$ do not need to be considered because of
axiom PGA3.
For eight of the remaining twelve forms, the equation to be proved
follows immediately from the equations to be proved for the other forms,
to wit $\ph \conc Q$, $P \conc \ph \conc Q$, $P \conc \ph\rep$ and
$P \conc (Q \conc \ph)\rep$, the axioms of \PGA, the defining equations
for thread extraction, and the easy to prove fact that
$\extr{P \conc \fjmp{0}} = \extr{P}$.

In the case of the form $\ph \conc Q$, the equation concerned is easily
proved by induction on $n$.
In the case of the form $P \conc \ph \conc Q$, only $P$ in which the
repetition operator does not occur need to be considered  because of
axiom PGA3.
For such $P$, the equation concerned is easily proved by induction on
the length of $P$, using the equation proved for the form $\ph \conc Q$.
In the case of the form $P \conc \ph\rep$, only $P$ in which the
repetition operator does not occur need to be considered  because of
axiom PGA3.
For such $P$, the equations for the approximating forms $P \conc \ph^k$
are easily proved by induction on $k$, using the equation proved for the
form $P \conc \ph \conc Q$.
From these equations, the equation for the form $P \conc \ph\rep$
follows using \AIP.
In the case of the form $P \conc (Q \conc \ph)\rep$, the equation
concerned is proved like in the case of the form $P \conc \ph\rep$.
\qed
\end{proof}

The following theorem states rigorously that the projections $\pgagpga$
and $\bpgagpga{k}$ give rise to instruction sequences with the same
behaviour.
\begin{theorem}
\label{theorem-labels}
For each closed \bPGAg{k} term $P$,
$\extr{\pgagpga(P)} = \extr{\bpgagpga{k}(P)}$.
\end{theorem}
\begin{proof}
Because
$\pgagpga(u_1 \conc \ldots \conc u_n) =
 \pgagpga(u_1 \conc \ldots \conc u_n \conc (\goto{1})\rep)$ and
$\bpgagpga{k}(u_1 \conc \ldots \conc u_n) =
 \bpgagpga{k}(u_1 \conc \ldots \conc u_n \conc (\goto{1})\rep)$,
we only  consider the case where the repetition operator occurs in $P$.

We make use of an auxiliary function $\extr{\ph,\ph}$.
This function determines, for each natural number and closed \PGA\ term
in which the repetition operator occurs, a closed term of \BTA\ with
guarded recursion.
The function $\extr{\ph,\ph}$ is defined as follows:
\begin{ldispl}
\begin{gceqns}
\extr{i,u_1 \conc \ldots \conc u_n \conc
        (u_{n+1} \conc \ldots \conc u_m)\rep}
\\ \hfill {} =
\extr{u_i \conc \ldots \conc u_m \conc
      (u_{n+1} \conc \ldots \conc u_m)\rep}
& \mif 1 \leq i \leq m\;,
\\
\extr{i,u_1 \conc \ldots \conc u_n \conc
        (u_{n+1} \conc \ldots \conc u_m)\rep} = \DeadEnd
& \mif \Not 1 \leq i \leq m\;.
\end{gceqns}
\end{ldispl}

Let
$P = u_1 \conc \ldots \conc u_n \conc
 (u_{n+1} \conc \ldots \conc u_m)\rep$ be a closed \bPGAg{k} term,
let $P' = \pgagpga(P)$, and let $P'' = \bpgagpga{k}(P).$
Moreover, let $\funct{\rho}{\Nat}{\Nat}$ be such that
$f(i) = (k + 3) \mul (i - 1) + 1$.
Then it follows easily from the definitions of $\extr{\ph,\ph}$,
$\extr{\ph}$, $\pgagpga$ and $\bpgagpga{k}$, the axioms of \PGA\ and
Proposition~\ref{prop-jump-chains} that for $1 \leq i \leq m$:
\pagebreak[2]
\begin{ldispl}
\begin{aceqns}
\extr{i,P'} & = & a \bapf \extr{i + 1,P'}
 & \mif u_i = a\;, \\
\extr{i,P'} & = & \pcc{\extr{i + 1,P'}}{a}{\extr{i + 2,P'}}
 & \mif u_i = \ptst{a}\;, \\
\extr{i,P'} & = & \pcc{\extr{i + 2,P'}}{a}{\extr{i + 1,P'}}
 & \mif u_i = \ntst{a}\;, \\
\extr{i,P'} & = & \extr{i + 1,P'}
 & \mif u_i = \lbl{l}\;, \\
\extr{i,P'} & = & \extr{i + n,P'}
 & \mif u_i = \goto{l} \And \tgt_i(l) = n\;, \\
\extr{i,P'} & = & \Stop
 & \mif u_i = \halt\;.
\end{aceqns}
\end{ldispl}
and
\begin{ldispl}
\begin{aceqns}
\extr{\rho(i),P''} & = & a \bapf \extr{\rho(i + 1),P''}
 & \mif u_i = a\;, \\
\extr{\rho(i),P''} & = &
\pcc{\extr{\rho(i + 1),P''}}{a}{\extr{\rho(i + 2),P''}}
 & \mif u_i = \ptst{a}\;, \\
\extr{\rho(i),P''} & = &
\pcc{\extr{\rho(i + 2),P''}}{a}{\extr{\rho(i + 1),P''}}
 & \mif u_i = \ntst{a}\;, \\
\extr{\rho(i),P''} & = & \extr{\rho(i + 1),P''}
 & \mif u_i = \lbl{l}\;, \\
\extr{\rho(i),P''} & = & \extr{\rho(i + n),P''}
 & \mif u_i = \goto{l} \And \tgt_i(l) = n\;, \\
\extr{\rho(i),P''} & = & \Stop
 & \mif u_i = \halt
\end{aceqns}
\end{ldispl}
(where $tgt_i$ is as in the definition of $\pgagpga$).
Because $\extr{\pgagpga(P)} = \extr{1,P'}$ and
$\extr{\bpgagpga{k}(P)} = \extr{\rho(1),P''}$, this means that
$\extr{\pgagpga(P)}$ and $\extr{\bpgagpga{k}(P)}$ are solutions of the
same guarded recursive specification.
Because guarded recursive specifications have unique solutions, it
follows that $\extr{\pgagpga(P)} = \extr{\bpgagpga{k}(P)}$.
\qed
\end{proof}
The projection $\bpgagpga{k}(P)$ yields only closed \PGA\ terms that
do not contain jump instructions $\fjmp{l}$ with $l > k + 3$.
Hence, we have the following corollary of Theorem~\ref{theorem-labels}:
\begin{corollary}
\label{corollary-labels-1}
For each closed \bPGAg{k} term $P$, there exists a closed \PGA\ term $P'$
not containing jump instructions $\fjmp{l}$ with $l > k + 3$ such that
$\extr{\pgagpga(P)} = \extr{P'}$.
\end{corollary}
It follows from Corollary~\ref{corollary-labels-1} that, if a regular
thread cannot be denoted by a closed \PGA\ term that does not contain
jump instructions $\fjmp{l}$ with $l > k + 3$, it cannot be denoted by a
closed \bPGAg{k} term.
Moreover, it is known that, for each $k \in \Nat$, there exists a closed
\PGA\ term for which there does not exist a closed \PGA\ term not
containing jump instructions $\fjmp{l}$ with $l > k + 3$ that denotes
the same thread (see e.g.~\cite{PZ06a}, Proposition~3).
Hence, we also have the following corollary:
\begin{corollary}
\label{corollary-labels-2}
For each $k \in \Nat$, there exists a closed \PGA\ term $P$ for which
there does not exist a closed \bPGAg{k} term $P'$ such that
$\extr{P} = \extr{\pgagpga(P')}$.
\end{corollary}

\section{Conclusions}
\label{sect-concl}

Program algebra is a setting suited for investigating single-pass
instruction sequences.
In this setting, we have shown that each behaviour that can be produced
by a single-pass instruction sequence under execution can be produced by
a single-pass instruction sequence without jump instructions if use can
be made of Boolean register services.
This is considered an interesting expressiveness result.
An important variant of program algebra is obtained by leaving out
jump instructions and adding labels and goto instructions.
We have also shown that a bound to the number of labels restricts
the expressiveness of this variant.
Earlier expressiveness results on single-pass instruction sequences as
considered in program algebra are collected in~\cite{PZ06a}.

Program algebra does not provide a notation for programs that is
intended for actual programming.
However, to demonstrate that single-pass instruction sequences as
considered in program algebra are suited for explaining programs in the
form of assembly programs as well as programs in the form of structured
programs, a hierarchy of program notations rooted in program algebra is
introduced in~\cite{BL02a}.
One program notation belonging to this hierarchy, called \PGLDg, is a
simple program notation, close to existing assembly languages, with
labels and goto instructions.
We remark that a projection from the set of all \PGLDg\ programs to the
set of all closed \PGAg\ terms can easily be devised.

The idea that programs are in essence single-pass instruction sequences
underlies the choice for the name program algebra.
The name seems to imply that program algebra is suited for investigating
programs in general.
We do not intend to claim this generality, which in any case does not
matter when investigating single-pass instruction sequences.
The name program algebra might as well be used as a collective name for
algebras that are based on any viewpoint concerning programs.
To our knowledge, it is not common to use the name as such.

\subsection*{Acknowledgements}
We thank Alban Ponse, colleague at the University of Amsterdam, and
Stephan Schroevers, graduate student at the University of Amsterdam, for
carefully reading a preliminary version of this paper and pointing out
some flaws in it.

\bibliographystyle{spmpsci}
\bibliography{TA}


\end{document}